\documentclass[10pt,letterpaper]{article}
\usepackage[top=0.85in,footskip=0.75in,marginparwidth=1in]{geometry}

\usepackage{amsmath}
\usepackage{amssymb}
\usepackage[bookmarks,        % add hyperlinks to the document
            colorlinks,
            plainpages=false]{hyperref}                    
\usepackage{lmodern}          % use the Latin Modern Font
\usepackage[T1]{fontenc}      % understand 8 bit fonts
\usepackage[latin1]{inputenc} % understand 8 bit latin1 input
\usepackage[english]{babel}   % use german hyphenation
\usepackage{graphicx}         % learn how to include graphics
\usepackage{calc}             % teach latex to calculate
\usepackage{makeidx}          % lets have an index
\makeindex                    % and enable the index
\usepackage{verbatim}         % a better verbatim environment

\graphicspath{{./}{figs/}}

\begin{document}

\begin{flushleft}
{\Large
\textbf\newline{In reply to Faes et al. and Barnett et al. regarding ``A study of problems encountered in Granger causality analysis from a neuroscience perspective''}
}
\newline
% authors go here:
\\ 
Patrick A. Stokes\textsuperscript{1},
Patrick L. Purdon\textsuperscript{1,*}
\\
\bigskip
1. Department of
    Anesthesia, Critical Care and Pain Medicine, Massachusetts General
    Hospital, 55 Fruit Street, Gray-Bigelow 444, Boston, MA 02144
\\
\bigskip
* patrickp@nmr.mgh.harvard.edu

\end{flushleft}

We would like to thank the authors Faes et al. \cite{Faes2017}, as well as Barnett et al. \cite{Barnett2017}, for their thoughtful commentary on our paper \cite{Stokes2017}. The main points of our work were to 1) characterize statistical properties of the traditional computation of Granger-Geweke (GG) causality, and 2) to analyze how the dynamics of the system are represented in the GG-causality measure.

Barnett et al. \cite{Barnett2017} and Faes et al. \cite{Faes2017} both point out that the issues with bias and variance in the conditional GG-causality can be addressed using a state space approach and a single model fit.  This is clearly the case, and is demonstrated in the simulation studies in Faes et al. \cite{Faes2017}. We regret that we were unaware of the earlier papers on GG-causality using state space models by Barnett and Seth \cite{Barnett2015} and Solo \cite{Solo2016}.  The original submission of our paper was in October of 2014, and a considerable time elapsed before our re-submitted manuscript was ultimately accepted, and our coverage of the literature during that span of time was not up to date.  We also described the state space solution to these problems in Dr. Stokes' Ph.D. thesis \cite{Stokes} in January 2015, but felt it was important to first characterize and describe the problem, before laying out a solution to that problem.  Along those lines, we believe our paper makes an important contribution by illustrating the form of the bias and variance, particularly in frequency domain, when separate model fits are employed.  As Faes et al. \cite{Faes2017} suggest, and as we have noted in other discussions surrounding our paper, many analysts continue to use separate model fits while performing GG-causality analysis, likely unaware of the problems with this approach.  Again, these computational issues can be avoided by using a single model fit, preferably using the state space approach \cite{Barnett2015, Solo2016, Faes2017, Stokes}. \footnote{We also acknowledge that the MVGC toolbox uses a single model fit as described in \cite{Barnett2014}, albeit with a different approach to spectral factorization. We regret our statement suggesting otherwise in our original manuscript.}	

Barnett et al. \cite{Barnett2017} emphasize that Granger causality reflects a ``directed information flow.'' But how does one meaningfully interpret that information flow?  Our analysis suggests that the receiver independence property is highly problematic for neuroscience studies, where the objective is typically to identify and/or characterize the mechanism of some observed effect. As we have shown, the dynamics of the effect nodes are absent in GG-causality. Oscillations play an important role in systems neuroscience, and interpretation of causality measures appears particularly problematic in systems with strong frequency-dependent structure.  Studies of oscillatory phenomena are invariably geared towards understanding the factors that contribute to oscillations observed at specific frequencies.  Ignoring these observed dynamics is simply not compatible with the goal of understanding them. 

As Barnett et al. \cite{Barnett2017} state, one can make a distinction between physiological or ``physical causal mechanisms'' and ``directed information flow.'' However, we perceive that in practice, the need to interpret and ascribe meaning to data analyses would tend to lead investigators to interpret ``directed information flow'' in mechanistic terms. So the notions of ``information flows'' versus mechanisms, though distinct in the abstract, might not be distinguished in practice.

While GG-causality is decipherable in reference to the selected model and its component dynamics, it is not understandable without these details. Unfortunately, many GG-causality works do not provide the estimated model, much less a breakdown of its component dynamics. More fundamentally, treating the causality as ``the result,'' overlooks that it is a statement about the chosen model and the product of the preceding modeling process. If a different model is chosen, then the causality may obviously change; and if the model is inadequate for the data or question, then any inference will be worthless.   

We were careful to focus the analysis in our paper on GG-causality. In the discussion section, we also expressed concerns that other causality  measures with distinct formulations and properties might have their own interpretational problems, complicated further by the fact that these methods are often treated interchangeably or referred to collectively as ``Granger causality.''  But we did not intend to dismiss efforts to develop improved methods for analyzing directed dynamical influences.   To the contrary, we believe that such methods will be essential for gaining meaningful insights from modern neuroscience data.  As we try to emphasize in our paper \cite{Stokes2017}, a crucial priority will be to ensure that the models and derived quantities correspond appropriately to the scientific questions of interest. Developing such methods will require a closer partnership between neuroscientists and quantitative scientists going forward. In the meantime, as we suggest in our paper, a good starting point would be for analysts to pay more attention to the underlying models, the dynamics they represent, and the overall modeling process, all of which form the foundations for subsequent inferences on directed influences.

%This defines the bibliographies style. Search online for a list of available styles.
%\bibliographystyle{unsrt} % {abbrv} % {pnas-new} % 

%This is where your bibliography is generated. Make sure that your .bib file is actually called library.bib
%\bibliography{biblio}

\begin{thebibliography}{1}

\bibitem{Faes2017}
Luca Faes, Sebastiano Stramaglia, and Daniele Marinazzo.
\newblock On the interpretability and computational reliability of
  frequency-domain {G}ranger causality.
\newblock {\em arXiv}, stat.ME, Aug 2017.

\bibitem{Barnett2017}
Lionel Barnett, Adam~B Barrett, and Anil~K Seth.
\newblock Reply to {S}tokes and {P}urdon: A study of problems encountered in
  {G}ranger causality analysis from a neuroscience perspective.
\newblock {\em arXiv}, stat.ME, Aug 2017.

\bibitem{Stokes2017}
Patrick~A Stokes and Patrick~L Purdon.
\newblock A study of problems encountered in {G}ranger causality analysis from
  a neuroscience perspective.
\newblock {\em Proc Natl Acad Sci USA}, 114(34):E7063--E7072, Aug 2017.

\bibitem{Barnett2015}
Lionel Barnett and Anil~K Seth.
\newblock {G}ranger causality for state-space models.
\newblock {\em Phys. Rev. E}, 91(4):040101, Apr 2015.

\bibitem{Solo2016}
Victor Solo.
\newblock State-space analysis of {G}ranger-{G}eweke causality measures with
  application to f{MRI}.
\newblock {\em Neural Comput}, 28(5):914--49, May 2016.

\bibitem{Stokes}
Patrick~A Stokes.
\newblock {\em Fundamental Problems in {G}ranger Causality Analysis of
  Neuroscience Data}.
\newblock PhD thesis, Massachusetts Institute of Technology, 2015.

\bibitem{Barnett2014}
Lionel Barnett and Anil~K. Seth.
\newblock The {MVGC} multivariate {G}ranger causality toolbox: A new approach
  to {G}ranger-causal inference.
\newblock {\em Journal of Neuroscience Methods}, 223:50--68, Feb 2014.

\end{thebibliography}

\end{document}